\documentclass[english,prl,reprint]{revtex4-1}
\usepackage[T1]{fontenc}
\usepackage[latin9]{inputenc}
\usepackage{booktabs}
\usepackage{graphicx}
\usepackage{babel}

\begin{document}

\title{Efficacy of multimedia learning modules as preparation for lecture-based
tutorials in electromagnetism}

\author{J. Christopher Moore}

\affiliation{Department of Chemistry and Physics, Coastal Carolina University,
Conway, South Carolina, 29528, USA}
\email{moorejc@coastal.edu}

\begin{abstract}
We have investigated the efficacy of online, multimedia learning modules (MLMs) as preparation for in-class, lecture-based tutorials in electromagnetism
in a physics course for natural science majors (biology and marine science). Specifically, we report the results of a multiple-group pre/post-test research design comparing two groups receiving different treatments with respect to activities preceding participation in \emph{Tutorials in Introductory Physics}. The different pre-tutorial activities where as follows: (1) students were assigned reading from a traditional textbook, followed by a traditional lecture; and (2) students completed online multimedia learning modules developed by the Physics Education Research Group at the University of Illinois at Urbana Champaign (UIUC),
and commercially known as smartPhysics. The MLM treatment group earned significantly higher mid-term examination scores and larger gains in content knowledge as measured by the Conceptual Survey of Electricity and Magnetism (CSEM). Student attitudes towards "reformed" instruction in the form of active-engagement tutorials where also improved. Specifically, post-course surveys showed MLM-group students believed class time was more effective and the instructor was more clear than reported by non-MLM students, even though there was no significant difference between groups with respect to in-class activities and the same instructor taught both groups. MLM activities can be a highly effective tool for some student populations, especially when student preparation
and buy-in are important for realizing significant gains.
\end{abstract}
\maketitle

\section{I. Introduction}

Students perceive little value in reading the traditional physics textbook before material is introduced in the classroom \cite{podolefsky06}. As a result, it is not surprising that few students prepare for class in this way, even when specifically required to do so by the instructor \cite{cummings02,stelzer09}. Effective reformed pedagogies in physics often require some degree of student preparation pre-class. In particular, the concept of the "flipped" classroom requires significant time-on-task outside traditional class meetings \cite{platt00}. Care must be taken when implementing reformed pedagogies that rely on student preparation, specifically with respect to motivation, assessment of those activities, and assignment of value to individual methodologies.

The concept of replacing traditional textbook reading with online, multimedia learning modules (MLMs) for content has been discussed in the literature as a means to encourage student pre-class preparation. For example, improved student learning and attitudes towards instruction have been found when reading assignments have been replaced with online multimedia videos in physics classes across various institutions \cite{chen10,stelzer10,sadaghianiTPT12}. In particular, it has been shown that students arrive to class better prepared for learning when completing MLMs versus traditional reading assignments for large-enrollment lecture-based courses and hybrid-online formats \cite{chen10,sadaghiani11,sadaghiani12}.

There has been recent interest in combining the concept of MLM pre-instruction activities with studio-style in-class pedagogies, such as \emph{Workshop Physics} and/or \emph{Tutorials in Introductory Physics} (TIP) \cite{laws97,laws91,slezak11,finkelstein05,cruz10,mcdermott02}. In this study, we are particularly interested in the effectiveness of the inverted classroom, with MLMs replacing the lecture outside of class time and TIP dominating the face-to-face instructional time. This type of implementation would be particularly useful at small- to mid-sized institutions without graduate teaching assistants and no defined recitation sections. 

However, successful implementation of new pedagogies requires caution, specifically because research shows that different populations can respond very differently to reformed instruction based on scientific reasoning ability, motivation, major, and general academic preparation \cite{moorerubbo12,moore12,finkelstein05}. In particular, non-science majors in college level courses may respond differently than science majors to the same instruction. Even different responses across science majors may be possible, with natural science majors (biology, marine science, etc.) responding to pedagogies in completely different ways than physical science majors.

In this article, we describe MLM pre-class activities and present an implementation in the second-semester calculus-based introductory course for natural science majors that utilizes TIP. In particular, we compare students preparing for the tutorials through either a combination of textbook reading assignments and traditional lecture, or MLM activities. We first describe MLMs, our student populations, the instructional environment for student groups utilizing  MLMs and those not using them, and how we measured efficacy. We then describe the MLMs effectiveness by comparing student exam scores, their performance on a validated assessment, and on student affect via surveys.


\section{II. Background}

In this section, we discuss multimedia learning and the development and deployment of MLMs in instruction. We also discuss the specific physics-based MLMs used in the courses under study. The student population and course format is described, since different populations can respond very differently to reformed instruction, and factors such as the classroom environment and particular pedagogical approach can influence results of multiple-group studies. Finally, we discuss the specific study design and how measures of efficacy where obtained and analyzed.

\subsection{A. Multimedia learning modules}

MLMs as an instructional tool have been developed as a means towards reducing cognitive load in the learning process \cite{mousavi95}. By mixing auditory and visual presentations, meaningful learning can take place by helping the learner make connections between multiple representations of the same content without taxing limited capacity memory channels \cite{mayer97}. Although to some extent, multiple representations can be introduced in both lecture and the textbook, these tools are limited primarily to single modes of delivery. With respect to textbook reading, there is also significant practical concern about whether or not students are using it as a learning tool at all in physics courses \cite{cummings02,stelzer09}. Furthermore, there is a long history of research findings that show the traditional physics lecture is ineffective, with more recently developed classroom-based active engagement reformed pedagogies demonstrating significant gains in student learning. Interestingly, research in educational psychology has shown that multimedia approaches to the presentation of content outside of the classroom can result in the same type of cognitive processes associated with these active learning approaches in the classroom \cite{mayer01}. This suggests great promise for MLMs as a pre-class preparation for deeper learning. 

For this study, we have used online MLMs developed by the Physics Education Research Group at the University of Illinois at Urbana Champaign (UIUC),
and commercially known as smartPhysics \cite{smartphysics}. These MLMs are flash animations that are designed to introduce core concepts within a traditional physics course in relatively short presentations of approximately 15 minutes. The authors refer to these MLMs as "pre-lectures," since their intended purpose is as activities preceding the classroom lecture period on the same subject. Each pre-lecture includes a narration with animations, resulting in a highly visual and auditory experience where certain concepts can "come alive" in a way not reproducible on the traditional whiteboard.

\begin{figure}[]
\includegraphics[]{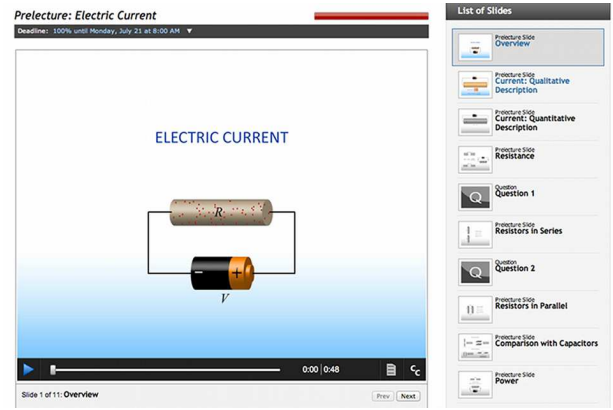}
\caption{Screen capture of a smartPhysics pre-lecture on Coulomb's Law.}
\label{prelecture}
\end{figure}

Figure \ref{prelecture} shows a screenshot of the student view for a smartPhysics MLM pre-lecture on electric current. The entire 15 minute pre-lecture is split into multiple 1-2 minute animated slides on smaller concepts, as shown on the right side of the figure. The student has access to a play/pause button for each section of the pre-lecture, as seen in the bottom-left corner of Figure \ref{prelecture}; however, a student cannot proceed to the next section until viewing the current section. After all sections have been viewed, the student is free to browse the content for review. Also,  two to three multiple choice questions for each pre-lecture are embedded in-between some of the animated slides, which can be seen on the right side of Figure \ref{prelecture} as dark slides emblazoned with a large letter Q. In order for students to proceed to the next slide, they must correctly answer the question. Feedback is automatic, and if they incorrectly answer then the student is led through a short tutorial designed to lead them to the correct answer. 

It should be pointed out that the design of the MLMs used in this study where informed by research in both physics education and multimedia learning. A more in-depth discussion of how the authors of the MLMs utilized the relevant literature in the design process is available in reference \cite{sadaghiani11}. Furthermore, the efficacy of the MLMs have been verified in multiple other settings \cite{chen10,stelzer10,sadaghianiTPT12,sadaghiani11,sadaghiani12}.

\begin{table*}
\caption{List of MLM pre-lecture topics used during the course \cite{smartphysics}.}
\begin{tabular}{lll}
\toprule 
Electrostatics & DC Circuits & Magnetism\tabularnewline
\midrule
Coulomb's Law & Conductors and Capacitors & Magnetism\tabularnewline
Electric Fields & Electric Current & Biot-Savart Law\tabularnewline
Electric Flux and Field Lines & Kirchhoff's Rules & Ampere's Law\tabularnewline
Gauss' Law &  & Motional EMF\tabularnewline
Electric Potential Energy &  & Faraday's Law\tabularnewline
Electric Potential &  & \tabularnewline
\bottomrule
\end{tabular}
\label{topics}
\end{table*}

\begin{table*}
\caption{List of tutorials completed in-class that are associated with the
pre-lectures in tab. 1. Tutorials used correspond exactly to those
found in \emph{Tutorial in Introductory Physics} by McDermott, et
al.\cite{mcdermott02}}
\begin{tabular}{lll}
\toprule 
Electrostatics & DC Circuits & Magnetism\tabularnewline
\midrule
Charge & A model for circuits Part 1 & Magnets and magnetic fields\tabularnewline
Electric field and flux & A model for circuits Part 2 & Magnetic interactions\tabularnewline
Gauss' law &  & Lenz' law\tabularnewline
Electric potential difference &  & Faraday's law and applications\tabularnewline
Capacitance &  & \tabularnewline
\bottomrule
\end{tabular}
\label{tutorials}
\end{table*}


\subsection{B. Course information and student population}
\label{course}

Coastal Carolina University (CCU) is a primarily-undergraduate, comprehensive university in South Carolina, USA. During the summer of 2011, a second-semester calculus-based physics course (PHYS 212) at CCU was selected to pilot an implementation of MLMs for most of the major content within the course. We served as a beta test site for the commercial smartPhysics product eventually put into production by W.H. Freeman \cite{smartphysics}. During the fall of 2011, two sections of the same course also implemented MLMs. All courses implementing MLM content where taught by the same primary instructor. In this paper, we compare student learning and affect for sections implementing MLMs with sections previously taught by the same instructor without MLMs.

PHYS 212 is a calculus-based physics course covering content such as fluids, waves, thermodynamics, electricity and magnetism. The population of this course is predominantly composed of natural science majors from the Departments of Biology and Marine Science. At CCU, all introductory physics courses are taught in a lecture room designed for the Student Centered Activities for Large Enrollment University Physics (SCALE-UP) model \cite{beichner99}. A SCALE- UP course incorporates the high-impact practice of collaborative assignments and projects by fusing lecture, laboratory, and recitation into a single entity. For PHYS 212, typically between two and three 24-seat sections are combined into one large classroom holding up to 72 students. Students work at round tables that seat six students each. The course is led by a primary instructor and usually has one or two instructors serving in backup roles, with at least one undergraduate Learning Assistant \cite{otero10}.

During face-to-face class time, a lecture-based implementation of \emph{Tutorials in Introductory Physics} was utilized for all groups, including the non-MLM group \cite{mcdermott02}. Similar to a traditional TIP implementation, students proceeded through TIP materials in groups, but with whole-class "checkouts" rather than instructor-intensive, individual group checkouts. We also intersperse short "micro-lectures" in between TIP activities, as well as Peer Instruction activities \cite{mazur97}. This adaptation of TIP does stray in some significant ways from the intentions of the curriculum developers. However, this adaptation is necessary for logistical reasons. This type of implementation would be particularly useful at similar small- to mid-sized institutions without graduate teaching assistants and no defined recitation sections. For both the MLM and non-MLM courses, we also utilized Just-In-Time teaching and online, instant feedback homework, which will be discussed in more detail in the next subsection.

Table \ref{topics} shows a list of the MLM pre-lecture topics assigned to students during the courses utilizing MLMs. Table \ref{tutorials} lists the TIP activities completed in face-to-face class meetings for both the MLM and non-MLM groups. For this study, we confine our analysis to the topics of traditional electricity and magnetism, such as electrostatics, direct current (DC) electric circuits, and magnetism. Other topics were discussed in the course as described above; however, measures of efficacy where collected only for the topics listed in Table \ref{topics}.

\subsection{C. Study Design}

\begin{figure}[]
\includegraphics[scale=0.3]{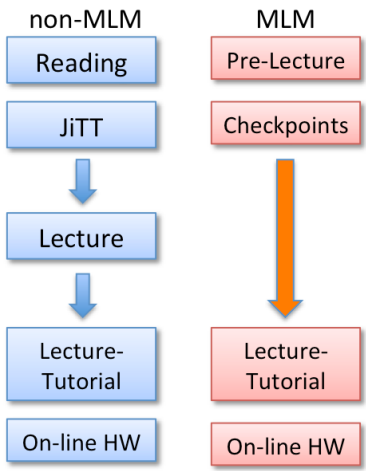}
\caption{Schematic diagram showing the study design.}
\label{studydesign}
\end{figure}

We conducted a multiple-group pre/post-test research design comparing two groups receiving different treatments with respect to activities preceding participation in TIP activities. Figure \ref{studydesign} summarizes the study design, where the different pre-tutorial activities where as follows: (1) students were assigned reading from a traditional textbook, followed by a traditional lecture; and (2) students completed online MLMs as described above. Four PHYS 212 sections served as our non-MLM group (N=58) and three sections of PHYS 212 served as our MLM treatment group (N=41). With respect to in-class activities, both groups had the same lead instructor and participated in the same lecture-tutorial activities. We should point out that the two groups were separated by different semesters, so they were not in the same class experiencing exactly the same in-class activities. Therefore, some variations in face-to-face instruction could have occurred, though this was kept to a minimum for the duration of the study.

An implementation of Just-in-Time Teaching (JiTT) was used with both groups \cite{novak99}. JiTT is a strategy that uses feedback between classroom activities and work that students do at home in preparation for the face-to-face meeting. Specifically, students complete some outside class activity such as reading or MLMs and then answer short, concept-based questions online before the beginning of the class. The faculty member then uses the student responses to the questions to tailor the in-class materials to the expressed weaknesses of the particular class. 

For the non-MLM group, we used JiTT concept questions from reference \cite{novak99} that corresponded to the material. For the MLM group, JiTT-style "checkpoints" are automatically incorporated into the smartPhysics product that was being utilized \cite{smartphysics}. There was variation in the JiTT/checkpoint questions asked of each group; however, both sets of questions are based on research in physics education. Also, since JiTT as a fundamental feature allows the instructor to tailor in-class course content for individual classes, there is the potential for some topics in one class to receive more focus than in another. This should be kept in mind when evaluating the results comparing the MLM and non-MLM groups. However, for the content topics discussed in this article, there was very little difference between the strengths and weaknesses of the two groups pre-tutorials, which led to little variation in in-class tutorial activity. Although it appeared that MLM students were better prepared than non-MLM students (we did not measure this directly), both groups generally expressed difficulties with the same content.

For the non-MLM group, an instructor-led passive lecture session lasting between 45-50 minutes was incorporated. This lecture served as a more traditional introduction to material and in many ways was similar to the pre-lectures viewed by the MLM group, with the main exceptions being the lack of multimedia content. This added lecture session resulted in greater time-on-task for students that did complete the reading within the non-MLM group compared to the MLM group. Less content outside of the area of electromagnetism was covered during the non-MLM semesters to compensate for the added instructional time.

An instant feedback homework system was used with both groups. For the non-MLM group, tutorial-style homework problems were assigned using the MasteringPhysics online homework system and the textbook "Physics for Scientists and Engineers" by Randall Knight \cite{masteringphysics,knight07}. For the MLM group, an online homework system is automatically incorporated into the smartPhysics product that was being utilized \cite{smartphysics}. The style of the homework questions was similar for both groups, though the exact problems where different.


\section{III. Results}

The efficacy of MLMs was determined by comparing student groups performance on a validated assessment and exam scores. Student affect was measured via surveys. Specifically, learning gains on content in electricity and magnetism was measured via pre/post-test scores on the Conceptual Survey of Electricity and Magnetism (CSEM) \cite{maloney01}. In-class summative assessments in the form of exams where held constant across both groups, and each group's performance was compared. Finally, student course evaluations were administered at the end of each course for both groups and compared.

\subsection{A. Student learning gains}

The CSEM is a 32-item multiple choice assessment designed to test student understanding of electricity and magnetism concepts covered in the average introductory physics course. For all groups, we administered the CSEM both before instruction and after instruction. In particular, students completed the CSEM initially during the first class meeting of the semester, and then again during the last class meeting of the semester. No course credit was assigned for completing the assessment.

Each student's pre-test score $S_{pre}$ and post-test score $S_{post}$ was used along with the maximum possible score $S_{max}$ to calculate individual normalized learning gain $g$. Normalized gain is the ratio of the actual assessment score gain to the maximum possible gain, as follows \cite{hake98}:

\begin{equation}
\label{gain}
g=\frac{S_{post} - S_{pre}}{S_{max} - S_{pre}}.
\end{equation}
To compare the results across groups, we averaged the individual normalized gains for members of the group. We only present data for students that completed both the pre- and post-test. Since the assessment was not a required and graded component of the course, we not surprisingly had fewer students complete both offerings compared to the numbers enrolled in the courses.

\begin{figure}[t]
\includegraphics[]{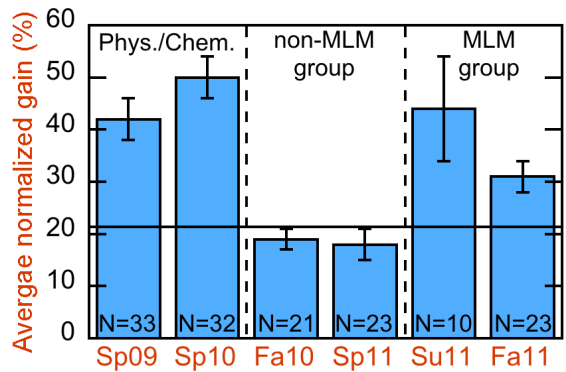}
\caption{Average normalized gain on the CSEM for six semesters of PHYS 212. Semesters Sp09 and Sp10 were taught to physics and chemistry majors at LU and did not incorporate MLM activities. Semesters Fa10 and Sp11 were taught to biology and marine science majors at CCU and did not incorporate MLM activities. Semesters Su11 and Fa11 were taught to biology and marine science majors at CCU and did incorporate MLM activities. In class tutorial activities were constant across all semesters. }
\label{csem}
\end{figure}

Figure \ref{csem} shows the average normalized gain on the CSEM for three different student groups across six different semesters of instruction. The solid horizontal line represents the national average normalized gain for the CSEM \cite{maloney01}. During the spring of 2009 (Sp09) and the spring of 2010 (Sp10), the same instructor taught a version of PHYS 212 at another institution, Longwood University (LU) in Farmville, VA, to a population of students consisting of primarily physics and chemistry majors. The exact same in-class instructional methodologies detailed in the previous section were utilized for these offerings as those during the fall 2010 (Fa10) and spring 2011 (Sp11) courses at CCU. Neither of these populations utilized MLMs. During the summer and fall or 2011 (Su11 and Fa11, respectively), MLMs were deployed.

An interesting and significant decrease in CSEM gains was observed from the spring to the fall of 2010. As discussed, successful implementation of new pedagogies requires caution, specifically because research shows that different populations can respond very differently to reformed instruction \cite{moorerubbo12,moore12,finkelstein05}. In this case, different responses across science majors was observed, with natural science majors (biology, marine science, etc.) responding to the same pedagogies in completely different ways than physical science majors. 

It should be pointed out that we have implemented an adaptation of TIP in both situations that strays in significant ways from the intentions of the curriculum designers. Therefore, the data should not be interpreted as condemnation of any particular pedagogy for any particular group. Our main point is certainly not that some specific pedagogy or collection of pedagogies fails to impact learning beyond the national average, but that any "un-tested" implementation could without proper concern for variations in population. In fact, it was this dramatic decrease in learning gains during 2010 that motivated this study. Anecdotal evidence suggested that the LU group utilized the text more before instruction than the CU group, leading to greater preparation during the tutorials. Motivation was also certainly a contributor, since self-selecting physics majors are generally more interested in learning physics than natural science majors.

Since our population in the LU group is significantly different from those at CU, we do not include them in our investigation of the efficacy of MLMs. However, our reason for including them in this report is to highlight that a successful adapted implementation of a pedagogy (lecture-based TIP at LU) can be dramatically altered when changing contexts, and that this could be caused by differences in preparations students make outside of the classroom.

\begin{table}
\caption{Average normalized learning gains on the CSEM for the non-MLM (N=44) and
MLM (N=33) groups.}
\begin{tabular}{ccc}
\toprule
\emph{<g>},\emph{ }non-MLM (\%) & \emph{<g>}, MLM (\%) & \emph{p}\tabularnewline
\midrule 
22$\pm$2 & 35$\pm$3 & <0.01\tabularnewline
\bottomrule
\end{tabular}
\label{csem_table}
\end{table}

The non-MLM group was made up of students from the Fa10 and Sp11 sections of PHYS 212, while the MLM group was composed of students in the Su11 and Fa11 sections. The average normalized gain on the CSEM for both groups is reported in Table \ref{csem_table}. The non-MLM group had an average normalized gain of $(22\pm2)$\%, which is statistically equivalent to the national average gain observed for courses not utilizing active engagement pedagogies (23\%) \cite{maloney01}. The MLM group had an average normalized gain of $(35\pm3)$\%. A two-sample location \emph{t}-test was used to determine whether or not the means of the two populations were equal. The MLM group had a significantly greater CSEM average learning gain compared to the non-MLM group valid at the \emph{p} < 1\% level. We found no statistical difference in average learning gains for the Fa10 and Sp11 groups, and the Su11 and Fa11 groups.

\subsection{B. Course examinations}

During all four semesters under study, students completed five closed-book exams, with four two-hour exams on content within specific learning units, and one two-hour cumulative exam after the end of the semester. All exams where composed of five free-response questions. Two questions were analytical problems similar to assigned homework problems, and three questions were free-response concept questions either taken directly or slightly modified from the sample exam questions in the Instructor's Guide to \emph{Tutorials in Introductory Physics} \cite{mcdermott02}. 

One exam focused on concepts in electrostatics, and another exam focused on topics in electromagnetism and DC circuits, consistent with the content described in Table \ref{topics}. These exams were the summative assessment of learning for these units, and for all four semesters studied (Fa10, Sp11, Su11, and Fa11), these two exams were the last two in-semester exams assigned. The same course instructor graded all of the exams. The exams for the Sp09 and Su11 semesters were identical. Likewise, the exams for the Sp11 and Fa11 semesters were identical, though different than the other two semesters. Reusing exams during non-consecutive semesters helped prevent old exams from being distributed while allowing for a common assessment across both groups.

\begin{figure}[t]
\includegraphics[scale=0.8]{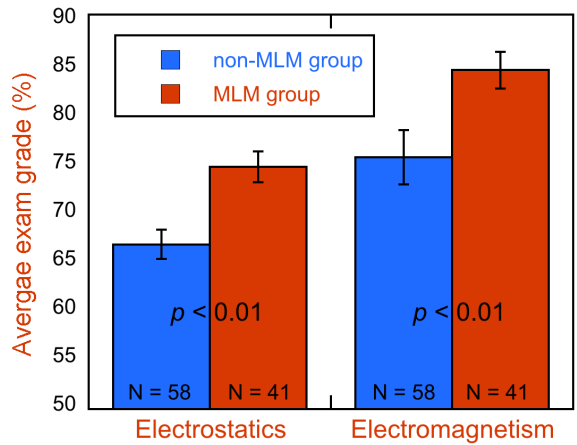}
\caption{Group average examination grades for two mid-term examinations: (1)
electrostatics, and (2) electromagentism. The same exams where used
in Fall 2011 and Summer 2011, and the same exams where used in Spring
2011 and Fall 2011.}
\label{exams}
\end{figure}

Figure \ref{exams} shows the average exam grade for the MLM and non-MLM groups for exams on electrostatics and electromagnetism. For the electrostatics exam, the non-MLM group had an average of $(67\pm2)\%$ while the MLM group had an average of $(75\pm2)\%$. For the electromagnetism exam, the non-MLM group had an average of $(76\pm3)\%$ while the MLM group had an average of $(85\pm2)\%$.  A one-tail unparied \emph{t}-test was used to determine whether or not the means of the two populations were equal. The MLM group had a significantly greater exam average for both exams compared to the non-MLM group valid at the \emph{p} < 1\% level. We found no statistical difference between exam grades when comparing between the two semesters of MLM and non-MLM instruction, which suggests the use of two different exams had little effect with respect to the average exam grade.

On both exams, the MLM group scored between 8 and 9\% higher than the non-MLM group. Both groups performed better on the electromagnetism exam than the electrostatics exam, which is consistent with observations from prior semesters. This is not well understood, but could be due to students' poor performance with Gauss' Law compared to surprisingly consistent success with applications of the right-hand-rule in magnetism. This is also evident in the CSEM scores, where students score approximately 5-8\% higher on the magnetism questions compared to the electrostatic questions.

The overall increase in exam scores with MLMs is consistent with the similarly observed increase in CSEM learning gains. A similar increase in exam scores as a result of MLM use was observed for mechanics and electromagnetism content in other studies \cite{sadaghiani12,stelzer10}. This is the first study of MLM efficacy showing a consistent link between increasing gains in content knowledge as measured by a nationally validated instrument and performance on in-class examinations.

\subsection{C. Student perceptions of the instruction}

A survey on student attitudes towards the instructor was administered at the end of the semester for all courses discussed in this study. The survey consisted of seven questions with all questions answered via a 7-point Likert scale ranging from "Strongly Disagree" to "Strongly Agree." The survey used was administered during the last week of all classes and was part of CCU's standardized faculty evaluations used in all courses at the university.

\begin{table*}
\caption{Survey questions for the four topics of interest.}
\begin{tabular}{lll}
\toprule 
\emph{Clarity} & The instructor presented material in a clear and understandable way.\tabularnewline
\emph{Atmosphere} & The instructor established a comfortable learning atmosphere in the classroom.\tabularnewline
\emph{Class Time} & The instructor made good use of class time.\tabularnewline
\emph{Effectiveness} & The instructor was an effective teacher.\tabularnewline
\bottomrule
\end{tabular}
\label{survey}
\end{table*}

For this study, we were interested in the students' attitudes with respect to the following areas: (1) instructor clarity, (2) class atmosphere, (3) effective use of class time, and (4) instructor effectiveness. Table \ref{survey} shows the four questions on the survey used to elicit a response on these four areas. The other three questions asked students to assess the instructor on preparation for class, knowledge of subject matter, and enjoyment of teaching. These areas are not relevant to this study, and are therefore not discussed. It should be mentioned that the survey was designed as a student evaluation of the instructor, and not an assessment of the course. There were also no specific questions concerning the MLMs themselves. Student attitudes concerning specific course components, including MLMs, have been described elsewhere in the literature \cite{sadaghiani12}.

Responses to survey questions were scored on a 7-point scale and the average for each class was normalized to a percentage scale for comparison across groups. For example, a response of "Strongly Disagree" would be scored as a 1, and "Strongly Agree" would be scored as a 7. An average score of 5 would be normalized to 71.4\%.

Table \ref{evals_table} shows the average normalized scores on end-of-semester evaluations for all categories for both the non-MLM and MLM groups. With respect to instructor clarity, an average score of $(80\pm3)\%$ was reported by the non-MLM group, and a score of $(91\pm2)\%$ was reported by the MLM group. For class atmosphere, an average score of $(88\pm3)\%$ was reported by the non-MLM group, and a score of $(99\pm1)\%$ was reported by the MLM group. For effective use of class time, an average score of $(77\pm4)\%$ was reported by the non-MLM group, and a score of $(97\pm1)\%$ was reported by the MLM group. For instructor effectiveness, an average score of $(84\pm3)\%$ was reported by the non-MLM group, and a score of $(93\pm2)\%$ was reported by the MLM group. 

A one-tail unparied \emph{t}-test indicates a significant difference at the \emph{p} < 1\% level between the MLM and non-MLM groups with respect to students' attitudes in all areas except instructor effectiveness ($p=0.012$). There was no significant difference in any area between the Fa10 and Sp11 non-MLM groups, or the Su11 and Fa11 MLM groups.

The largest improvement in student attitudes towards the instructor was in the area of the effective use of class time, where a 20\% increase was observed. Also of interest is the large 11\% increase in instructor clarity. In particular, improvements in student attitudes concerning the instructor are interesting considering all classes had the same instructor and there was no significant difference in face-to-face activities, with the exception being an instructor-led passive lecture session for the non-MLM group.

\begin{table}
\caption{Normalized scores on end-of-semester student evaluations. For the
non-MLM group \emph{n} = 42, and for the MLM group \emph{n }= 37.
Only student perceptions of effectiveness where not significant at
the 1\% level.}
\begin{tabular}{cccc}
\toprule
 & non-MLM (\%) & MLM (\%) & \emph{p}\tabularnewline
\midrule
Clarity & 80$\pm$3 & 91$\pm$2 & <0.01\tabularnewline
Atmosphere & 88$\pm$3 & 99$\pm$1 & <0.01\tabularnewline
Class time & 77$\pm$4 & 97$\pm$2 & <0.01\tabularnewline
Effectiveness & 84$\pm$3 & 93$\pm$2 & 0.012\tabularnewline
\bottomrule
\end{tabular}
\label{evals_table}
\end{table}


\section{IV. Discussion}

The purpose of this study was to determine if MLMs could be effective for physics courses predominantly composed of natural science majors and that utilize a tutorials-style pedagogy for in-class meetings. The results suggest that MLMs were effective at improving student learning, as measured via a nationally validated instrument and examination scores. In particular, average student learning gains on the CSEM were 13\% higher and exam grades were between 8 and 9\% higher for students completing MLMs. We have also found that student attitudes towards the instructor of the course were improved. This result is non-intuitive, since the same instructor and similar face-to-face strategies were employed in all courses.

Students in the MLM group believed that the instructor presented the material in a clear and understandable way 11\% more than students in the non-MLM group. This result was initially surprising, since for the MLM group the in-class instructor/student interaction was almost completely in the context of TIP activities. These activities can be difficult and sometimes initially very confusing for students, exactly because they are designed to challenge previously held conceptions about physics content. For the non-MLM group, the instructor was spending considerably more time interacting with students and discussing the same content as the MLM group.

The improvement in student attitudes concerning instructor clarity may be attributable to the instructor shifting into the role of "clarifying agent" for the MLM group. For the non-MLM group, ideally the textbook is the student's initial introduction to the content in the course. However, as shown in previous studies, most students do not complete reading assignments, even when rewarding schemes through grading are implemented \cite{cummings02}. For the non-MLM group, this means that the instructor serves as the student's first introduction to course content during the lecture period.

Students have very poor pre-instruction understanding of the content in electromagnetism, especially compared to topics in mechanics \cite{maloney01}. They often arrive with very few conceptions about the topic at all, much less misconceptions. This means that for this set of topics, it is not surprising that their initial exposure to the material would be confusing. For the MLM group, a third-party was responsible for introducing the content in the form of MLMs. During the face-to-face class time, the instructor was able to serve as the clarifying agent. Although this cycle was also present with the non-MLM group, they lacked a third-party to "blame" for the initial confusion.

We did not measure student attitudes towards the MLMs themselves. However, students often complained in the moment about how confusing the MLMs were and how they learned more during class. This is consistent with studies showing MLMs ranking below lecture and interactive classroom activities in students' views about the usefulness of various course elements \cite{sadaghiani11}. The MLMs serve a critical role in preparing students for learning by introducing material and helping them understand what they do and do not understand. However, students do not necessarily realize this until after the course, or never at all.

The largest improvement in student attitudes towards the instructor was in the area of class time effectiveness. Students in the MLM group believed the instructor made good use of class time 20\% more than students in the non-MLM group. Considering the MLM group spent less in-class time on the same amount of material, this is not necessarily surprising; however, the students in the MLM group did not necessarily know what students in previous classes had experienced.   This improvement in student attitudes could be attributed to the same process described for instructor clarity. If students perceived the in-class activities to be clarifying agents, then they could have also perceived their time in class to be more useful than the non-MLM group. Although the non-MLM group worked on the same TIP activities in class, they also had to cope with much of their initial confusion in class during lecture-based introduction to the content.

It should be noted that the MLM groups were taught nearly one year after the non-MLM groups. This could have also provided the instructor more experience working with the population of students enrolled at CCU in natural science programs. This added experience could have resulted in improved micro-interactions with individual students that were not necessarily perceived consciously by the instructor. This should be taken into consideration when evaluating the results of this study, along with the variations in JiTT and homework activities. However, the reported improvements where achieved using less in-class time compared to the more traditional lecture-then-recitation approach to TIP employed with the non-MLM group. Even if learning gains had remained unchanged, the results would still suggest that MLMs are an effective replacement for the lecture in an inverted classroom strategy.

In summary, we have investigated the efficacy of online MLMs as preparation for in-class, lecture-based tutorials in electromagnetism in a physics course for natural science majors. Students utilizing MLMs demonstrated larger gains in learning, higher examination scores, and improved attitudes towards the instructor.


\section{acknowledgments}
The author would like to thank Freeman Worth Publishers for providing CCU students with free access to the smartPhysics website as part of a beta test. I would also like to thank Tim Stelzer and Mats Selen of the Physics Education Research Group at the University of Illinois at Urbana Champaign. They provided thoughtful discussion leading to the writing of this manuscript.

\bibliography{refs}

\end{document}